\documentclass[reqno]{amsart}
\usepackage{amssymb,graphics,graphicx,bbm,color}
\usepackage{multirow}

\definecolor{gray}{rgb}{0.93,0.93,0.93}
\definecolor{light-gold}{rgb}{0.99,0.97,0.78}

\def\be{\begin{equation}}
\def\ee{\end{equation}}
\def\bm{\begin{multline}}
\def\bfig{\begin{figure}[htb]}
\def\efig{\end{figure}}
\newcommand{\dd}{{\rm d}}
\newcommand{\e}[1]{\,{\rm e}^{#1}\,}
\newcommand{\ii}{{\rm i}}

\newcommand{\sumtwo}[2]{\sum_{\substack{#1 \\ #2}}}

\numberwithin{equation}{section}

\newtheorem{conjecture}{Conjecture}

\newcommand{\caC}{{\mathcal C}}
\newcommand{\caE}{{\mathcal E}}

\newcommand{\caL}{{\mathcal L}}

\newcommand{\bbC}{{\mathbb C}}
\newcommand{\bbE}{{\mathbb E}}
\newcommand{\bbN}{{\mathbb N}}
\newcommand{\bbP}{{\mathbb P}}

\newcommand{\bbZ}{{\mathbb Z}}

\makeatletter
  \def\tagform@#1{\maketag@@@{\footnotesize{(#1)}\@@italiccorr}}
\makeatother

\renewcommand{\eqref}[1]{(\ref{#1})}

\begin{document}


\title{A numerical study of the 3D random interchange and random loop models}

\author[A. Barp]{Alessandro Barp}
\address{Department of Physics, University of Warwick,
Coventry, CV4 7AL, United Kingdom}
\email{A.Barp@warwick.ac.uk}

\author[E.G. Barp]{Edoardo Gabriele Barp}
\address{Department of Physics, University of Warwick,
Coventry, CV4 7AL, United Kingdom}
\email{E.G.Barp@warwick.ac.uk}

\author[F.-X. Briol]{Fran\c cois-Xavier Briol}
\address{Department of Statistics, University of Oxford,
1 South Parks Road, Oxford OX1 3TG, United Kingdom}
\email{briol@stats.ox.ac.uk}

\author[D. Ueltschi]{Daniel Ueltschi}
\address{Department of Mathematics, University of Warwick,
Coventry, CV4 7AL, United Kingdom}
\email{daniel@ueltschi.org}

\subjclass{60K35, 82B20, 82B26}

\keywords{Random interchange model; random loop model; macroscopic loops; Poisson-Dirichlet distribution.}

\begin{abstract}
We have studied numerically the random interchange model and related loop models on the three-dimensional cubic lattice. We have determined the transition time for the occurrence of long loops. The joint distribution of the lengths of long loops is Poisson-Dirichlet with parameter 1 or $\frac12$.
\end{abstract}

\thanks{Work partially supported by a URSS grant from the University of Warwick.}
\thanks{\copyright{} 2015 by the authors. This paper may be reproduced, in its
entirety, for non-commercial purposes.}

\maketitle

\section{Introduction}
\label{sec intro}

The random interchange model is a stochastic process where transpositions are selected at random. The product of these transpositions gives a random permutation and the main question deals with its cycle structure. We consider variants where possible transpositions are restricted to nearest-neighbours of a regular cubic lattice. We also consider related loop models where ``crosses'' are replaced by ``double bars''. We provide evidence that a phase transition takes place where macroscopic loops occur. We give good estimates of the values of the parameters at the transition point, which should help to better comprehend the model. Finally, we compute moments of the lengths of the loops; it turns out that they are identical to those of the Poisson-Dirichlet distribution.

The random interchange model was invented by Harris \cite{Har}; T\'oth used it as a representation of the spin $\frac12$ quantum Heisenberg ferromagnet \cite{Toth}. Angel \cite{Ang} and Hammond \cite{Ham} obtained results for the model on trees. Schramm considered the variant on the complete graph and he proved that the joint distribution of the large cycle lengths is Poisson-Dirichlet of parameter 1, following a conjecture of Aldous \cite{Sch}; see also \cite{Ber} for some simplifications and extensions. Alon and Kozma have obtained remarkable identities that give the probability of cyclic permutations in terms of eigenvalues of the graph Laplacian, for arbitrary graphs \cite{AK}. These have allowed Berestycki and Kozma to obtain further results on the complete graph \cite{BK}.

A similar model with ``double bars'' instead of ``crosses'' was introduced by Aizenman and Nachtergaele in order to describe the spin $\frac12$ quantum Heisenberg model and the spin 1 model with biquadratic interactions \cite{AN}. It should be noticed that the representation of quantum systems involves the extra factor $\theta^{\rm \# loops}$ with $\theta=2,3$... In the survey \cite{GUW}, the authors conjectured that the joint distribution of the lengths of long loops, in dimensions three and higher, is given by the Poisson-Dirichlet distribution of parameter $\theta$. The two representations of T\'oth and Aizenman-Nachtergaele were recently combined so as to describe quantum models that interpolate between the two Heisenberg models, such as the spin $\frac12$ quantum XY model and further spin 1 models with SU(2)-invariant interactions \cite{Uel}. For these models, the joint distribution of long loops should be Poisson-Dirichlet with parameter $\theta/2$. The consequences of this structure have yet to be worked out. One such consequence is to identify the nature of symmetry breaking in the spin 1 model \cite{Uel2}.

The Poisson-Dirichlet distribution is conjectured to be a common feature of ``loop soups'' in dimensions three and higher. This was confirmed numerically in lattice permutations \cite{GLU} and in O(N) loop models \cite{NCSOS}. This was also confirmed, with a mathematically rigorous proof, in an annealed model of spatial permutations \cite{BU}. A numerical study of the spin 1 model \cite{VW} also provides indirect evidence, as explained in \cite{Uel2}. However, this conjecture is far from being accepted nowadays. It thus seems necessary to verify it also in those loop models that are related to quantum spin systems.

We introduce the random loop models in Section \ref{sec defs}. The conjectures about the joint distribution of the lengths of long loops are explained in Section \ref{sec conj}. Numerical evidence is presented in Section \ref{sec num} for the occurrence of a phase transition; we also discuss a comparison with bond percolation and quantum Heisenberg models. We investigate the presence of the Poisson-Dirichlet distribution in Section \ref{sec PD}.

\section{Random loop models}

\subsection{Definitions}
\label{sec defs}

Let $\Lambda = \{1,\dots,N\}^{3} \subset \bbZ^{3}$, and let $\caE_{\Lambda}$ denote the set of edges (nearest-neighbours) in $\Lambda$. Let $\beta>0$ and $u \in [0,1]$. To each edge of $\caE_{\Lambda}$ is associated an independent Poisson point process on the interval $[0,\beta]$ with two kinds of events:
\begin{itemize}
\item crosses occur with intensity $u$;
\item double bars occur with intensity $1-u$.
\end{itemize}
\begin{centering}
\bfig
\includegraphics{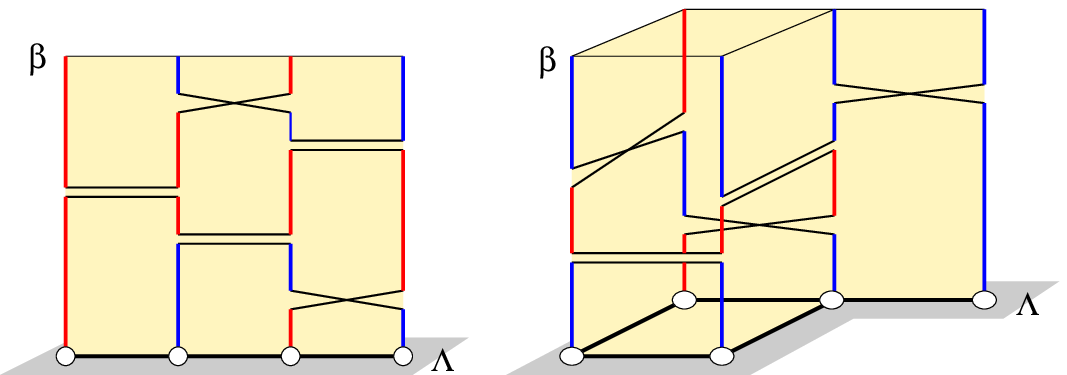}
\caption{Graphs and realisations of Poisson point processes, and their loops. In both cases, there are exactly two loops.}
\label{fig loops}
\efig
\end{centering}
See Fig.\ \ref{fig loops} for an illustration. Given a realisation $\omega$, we define loops by moving in the vertical direction, and by jumping to the neighbour whenever a cross or a double bar is encountered. If it is a cross, one continues in the same vertical direction; if it is a double bar, one continues in the opposite vertical direction. We assume periodic boundary conditions in the vertical direction. All trajectories must close, resulting in loops.
We use the following notation for the relevant random variables.
\begin{itemize}
\item $L_{1}(\omega), L_{2}(\omega), \dots$ denote the vertical lengths of all the loops in decreasing order (repeated with multiplicities); notice that $0 < L_{j}(\omega) \leq \beta |\Lambda|$ for all $j$.
\item $\ell_{1}(\omega), \ell_{2}(\omega), \dots$ denote the ``shadow lengths'' of the loops in decreasing order (repeated with multiplicities); that is, $\ell_j \in \{1,2,\dots,|\Lambda|\}$ is the number of sites at time 0 in the $j$th loop.
\end{itemize}
Notice that for all realisations $\omega$, we have $\sum_{j\geq1} L_{j}(\omega) = \beta |\Lambda|$ and $\sum_{j\geq1} \ell_{j}(\omega) = |\Lambda|$. The following are random partitions of $[0,1]$:
\be
\biggl( \frac{L_{1}(\omega)}{\beta |\Lambda|}, \frac{L_{2}(\omega)}{\beta |\Lambda|}, \dots \biggr), \qquad \biggl( \frac{\ell_{1}(\omega)}{|\Lambda|}, \frac{\ell_{2}(\omega)}{|\Lambda|}, \dots \biggr).
\ee

When $\beta$ is small, crosses and double bars are scarce and loops are small. But a phase transition occurs as $\beta$ grows and some loops become macroscopic. A few random partitions are displayed in Fig. \ref{fig partitions}; they were measured in a cube of volume $160^{3}$, for a value of $\beta$ that is above the critical parameter. Occurrence of macroscopic loops is manifest.

\bfig
\includegraphics[width=120mm]{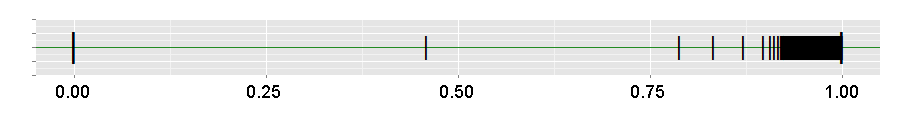}\\
\includegraphics[width=120mm]{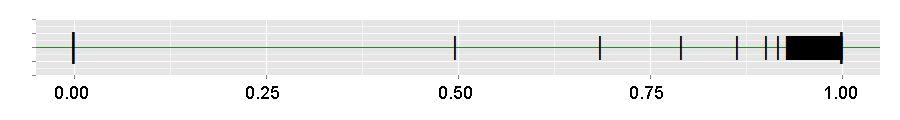}\\
\includegraphics[width=120mm]{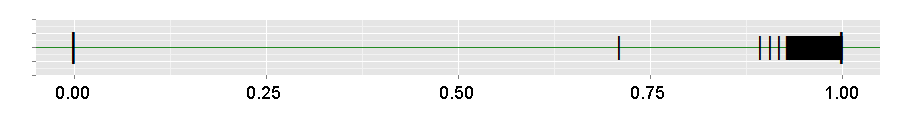}\\
\includegraphics[width=120mm]{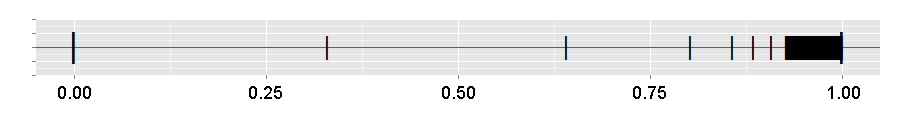}
\caption{Samples of random partitions observed in a cube of size $L = 160$, for $u=1$ and $\beta = 1$. Notice that $m(\beta)$ appears to be constant with approximate value 0.9.}
\label{fig partitions}
\efig

\subsection{Phase transition and universal behaviour}
\label{sec conj}

The patterns of the random partitions suggest a very interesting strong law of large numbers. It is worth describing in details, since it is expected to occur in all models of ``loop soups''.

\begin{conjecture}
\label{conj1}
There exists $m(\beta)$ such that, as $L\to\infty$, we have for almost all realisations $\omega$:
\be
\label{strong LLN}
\begin{split}
&\lim_{K\to\infty} \lim_{L\to\infty} \sum_{j=1}^{K} \frac{\ell_{j}(\omega)}{|\Lambda|} = m(\beta), \\
&\lim_{k\to\infty} \lim_{L\to\infty} \sum_{j: \ell_{j} \leq k} \frac{\ell_{j}(\omega)}{|\Lambda|} = 1 - m(\beta).
\end{split}
\ee
\end{conjecture}

The function $m(\beta)$ represents the mass of points in long loops. It is equal to 0 when $\beta$ is small, and it becomes positive when $\beta$ crosses the transition point. Eq.\ \eqref{strong LLN} says that loops are either microscopic (the length is of order 1) or macroscopic (the length is of order $|\Lambda|$); the number of sites that belong to loops of intermediate, mesoscopic length, has vanishing density.

Conjecture 1 is also expected to hold with $L_{j}/\beta$ instead of $\ell_{j}$, with the same $m(\beta)$.

The next conjecture is about the joint distribution of the lengths of the macroscopic loops, which should be Poisson-Dirichlet (PD). Let us recall the closely related Griffiths-Engen-McCloskey (GEM) distribution and its ``stick breaking'' construction. Let $X_{1}, X_{2}, \dots$ be i.i.d.\ Beta(1,$\vartheta$) random variables; their probability density function is $\vartheta (1-s)^{\vartheta-1}$ for $0 \leq s \leq 1$.
The following is a random sequence with GEM($\vartheta$) distribution:
\be
\label{GEM}
\bigl( X_{1}, (1-X_{1}) X_{2}, (1-X_{1}) (1-X_{2}) X_{3}, \dots \bigr).
\ee
It is not hard to verify that the sum of all these numbers is 1 with probability 1. Rearranging the numbers in decreasing order, we get a random partition with PD$_{[0,1]}(\vartheta)$ distribution. Multiplying each element by $m$, the distribution is PD$_{[0,m]}(\vartheta)$. We can formulate the conjecture about the joint distribution of macroscopic loops.

\begin{conjecture}
\label{conj2}
For any $k$, the joint distribution of $\ell_{1}, \dots, \ell_{k}$ converges to the joint distribution of the first $k$ elements of a random partition with PD$_{[0,m(\beta)]}(\vartheta)$ distribution, where
\[
\vartheta = \begin{cases} 1 & \text{if } u = 0 \text{ or } 1; \\ \frac12 & \text{if } u \in (0,1). \end{cases}
\]
\end{conjecture}

Here, $m(\beta)$ is the same quantity as in Conjecture \ref{conj1}. In random loop models with weights $\theta^{\# \text{loops}}$, the conjecture holds with $\vartheta = \theta$ if $u=0,1$, and $\vartheta=\theta/2$ if $u \in (0,1)$.

The heuristics for this conjecture goes back to Aldous' ideas for the complete graph, which Schramm eventually managed to turn into a proof \cite{Sch}. Its relevance for models with spatial structure was suggested in \cite{GUW,GLU}. In summary, the idea is to consider the stochastic process restricted on random partitions. Adding crosses or double bars result in splits or merges of elements of the partition. The rate at which two long loops merge may seem at first sight to depend on the exact geometry of the loops, which is very intricate. But averages take place when the loops are macroscopic, which is the case in dimension 3 and higher. The effective stochastic process is then a standard split-merge process, whose invariant measure is Poisson-Dirichlet.

The addition of a transition (cross or double bar) between different loops always results in a merge. When $u=1$, or when $u=0$ on a bipartite graph, the addition of a transition within the same loop always results in a split. It follows that the Poisson-Dirichlet distribution has parameter $\vartheta=1$. When $0<u<1$, the addition of a transition within the same loop may split it, or it may rewire it. The splits occur therefore at half the rate of the merges, and the Poisson-Dirichlet distribution has parameter $\vartheta=\frac12$. See \cite{GUW,GLU,NCSOS,Uel} for more detailed explanations.

\section{Phase transition and critical parameters}
\label{sec num}

\subsection{Fraction of sites in long loops}

We seek a convenient expression for the mass of sites in macroscopic loops $m(\beta)$. The expressions of Conjecture \ref{conj1} turn out to be inconvenient and we use Conjecture \ref{conj2} instead. It allows to relate $m(\beta)$ with the moments of the lengths of the loops, see Eq.\ \eqref{mbeta} below. We now give a derivation of this result.

We start with
\be
\label{le debut}
\begin{split}
\bbE \Bigl( \sum_{j\geq1} \Bigl( \frac{\ell_{j}}{|\Lambda|} \Bigr)^{2} \Bigr) &= \frac1{|\Lambda|^{2}} \sum_{j\geq1} \bbE \Bigl( \sum_{x,y \in \Lambda} 1_{(x,0) \in \text{$j$th loop}} \; 1_{(y,0) \in \text{$j$th loop}} \Bigr) \\
&= \frac1{|\Lambda|^{2}} \sum_{x,y \in \Lambda} \bbP \bigl( (x,0) \leftrightarrow (y,0) \bigr).
\end{split}
\ee
If we accept Conjecture \ref{conj2}, the probability that two sites $x,y$, that are far apart, belong to the same loop, is given by the probability that they both belong to long loops --- this is equal to $m(\beta)^{2}$ --- times the probability that two random numbers in $[0,1]$ belong to the same element in the partition. The latter probability can be calculated using the GEM random sequence \eqref{GEM}. The probability that both random numbers belong to the $j$th element is
\be
\begin{split}
\int_{0}^{1} \dd s_{1} \int_{0}^{1} \dd s_{2} \, \bbE_{{\rm GEM}(\vartheta)} &\bigl( 1_{s_{1}, s_{2} \in j \text {th element}} \bigr) = \bbE_{{\rm GEM}(\vartheta)}(Y_{j}^{2}) \\
&= \bbE_{{\rm Beta}(1,\vartheta)} \bigl( (1-X)^{2} \bigr)^{j-1} \, \bbE_{{\rm Beta}(1,\vartheta)} \bigl( X^{2} \bigr).
\end{split}
\ee
Elementary computations give $\bbE_{{\rm Beta}(1,\vartheta)} \bigl( (1-X)^{2} \bigr) = \frac\vartheta{\vartheta+2}$ and $\bbE_{{\rm Beta}(1,\vartheta)} \bigl( X^{2} \bigr) = \frac2{(\vartheta+1) (\vartheta+2)}$. The probability that two distant sites belong to the same loop is therefore approximately equal to
\be
\bbP \bigl( (x,0) \leftrightarrow (y,0) \bigr) = m(\beta)^{2} \sum_{j\geq1} \Bigl( \frac\vartheta{\vartheta+2} \Bigr)^{j-1} \frac2{(\vartheta+1) (\vartheta+2)} = \frac{m(\beta)^{2}}{\vartheta+1}.
\ee
Using \eqref{le debut}, we obtain an expression that is convenient for numerical calculations, namely
\be
\label{mbeta}
m(\beta) = \sqrt{(\vartheta+1) \, \bbE \Bigl( \sum_{j\geq1} \Bigl( \frac{\ell_{j}}{|\Lambda|} \Bigr)^{2} \Bigr)}.
\ee

Our numerical results are displayed in Fig.\ \ref{fig mbeta}. As expected, $m(\beta)$ is zero for small $\beta$ and positive for $\beta$ large enough. $m(\beta)$ is also continuous and increasing. Notice also that, when $u=1$, it converges to 1 as $\beta\to\infty$; indeed, all sites belong to long loops. It converges to a value smaller than 1 when $u=0$ or $\frac12$ because a density of small loops remains present in the system.

\bfig
\includegraphics[width=45mm]{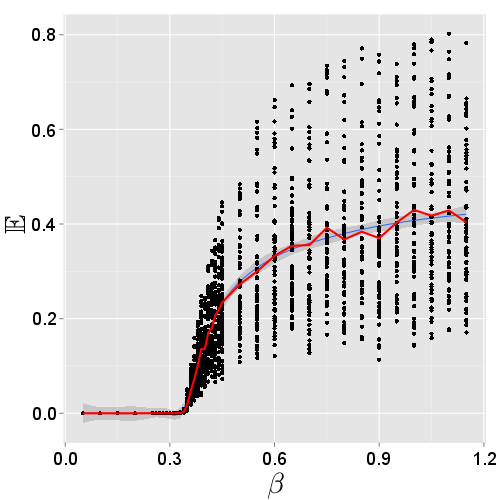}
\includegraphics[width=45mm]{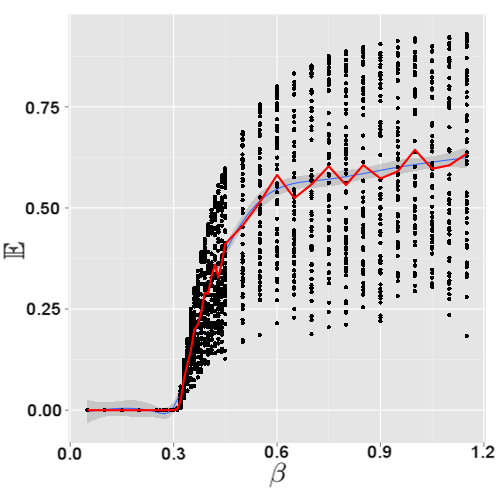}
\includegraphics[width=45mm]{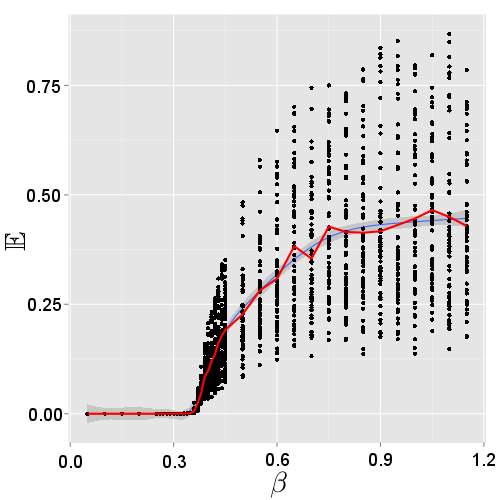}\\
\includegraphics[width=45mm]{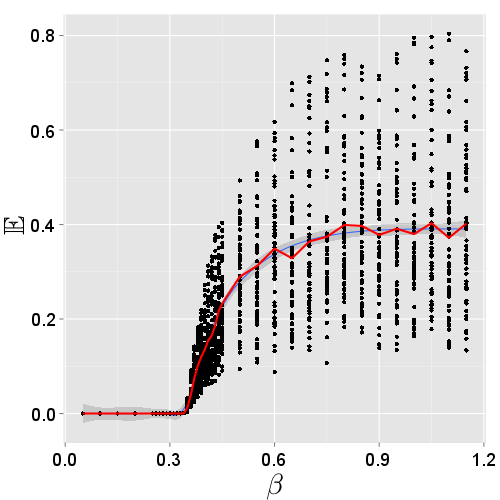}
\includegraphics[width=45mm]{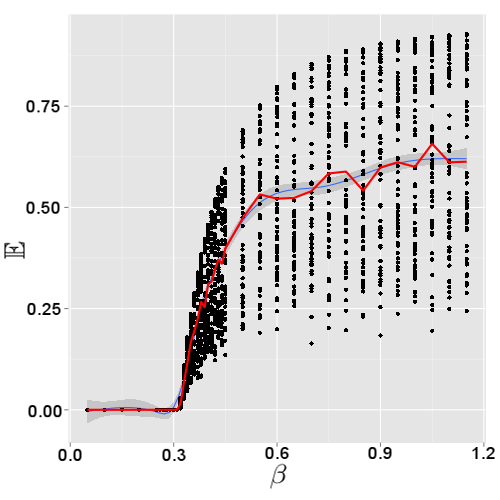}
\includegraphics[width=45mm]{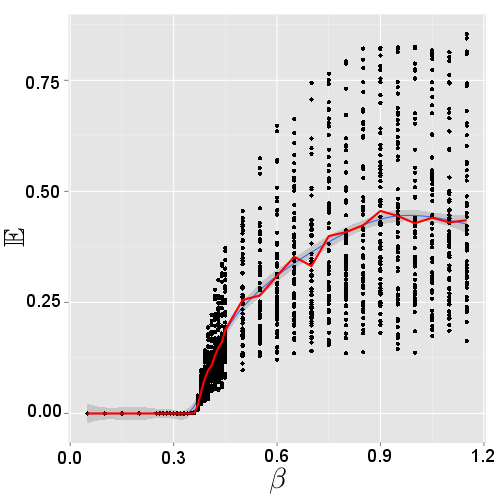}
\caption{Numerical values for $\bbE \bigl( \sum_{j\geq1} \bigl( \frac{\ell_{j}}{|\Lambda|} \bigr)^{2} \bigr)$ as function of $\beta$. The size of the cube is $L=80$ in the top row and $L=160$ in the bottom row. The parameter $u$ takes values 0 (left), 0.5 (centre), and 1 (right). The points represent the values of $\sum_{j} (\frac{\ell_{j}}{|\Lambda|})^{2}$ for many realisations; the curve gives the average.}
\label{fig mbeta}
\efig

\subsection{Value of the critical parameter $\beta_{\rm c}(u)$}

The previous results confirm the existence of a phase where $m(\beta)$ is positive. We define the critical parameter by
\be
\beta_{\rm c}(u) = \inf\{ \beta : m(\beta)>0 \}.
\ee
It is instructive to estimate its value and to investigate its dependence on the parameter $u$.
Our numerical results are depicted in Fig.\ \ref{fig betacrit}. We find that $\beta_{\rm c}(u)$ is a convex function of $u$. Its minimal value occurs when $u$ is close to 0.5. The derivative of $\beta_{\rm c}(u)$ diverges at $u=0$ and $u=1$. In retrospect, this is perhaps not surprising: The model with weight $2^{\# {\rm loops}}$ has more symmetry at $u=0$ and $u=1$,  SU(2) rather than U(1), so that a minor change in the value of $u$ has major consequences. This also explains why $\beta_{\rm c}(\frac12) < \beta_{\rm c}(0)$ and $\beta_{\rm c}(\frac12) < \beta_{\rm c}(1)$: Less symmetry means reduced fluctuations so that spontaneous magnetisation occurs more easily in systems with U(1) symmetry rather than SU(2). Somehow, this explanation should retain validity when the weight $2^{\# {\rm loops}}$ is not present.

\bfig
\includegraphics[width=90mm]{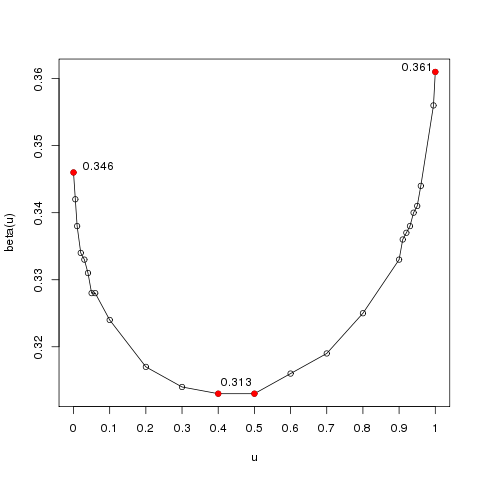}
\caption{Value of the critical parameter $\beta_{\rm c}$ as function of $u$.}
\label{fig betacrit}
\efig

\subsection{Comparison with bond percolation, Ising, and quantum Heisenberg models}

There are natural comparisons between the critical parameters of the model of random loops, of the Ising model (a.k.a.\ the $q=2$ random cluster model), and the quantum ferromagnetic and antiferromagnetic Heisenberg models.

\medskip\noindent
{\bf A. Bond percolation.}
Given a realisation $\omega$ of crosses and double bars on $\caE_{\Lambda} \times [0,\beta]$, there corresponds a percolation configuration $\eta = \eta(\omega)$ where $\eta_{xy} = 1$ if at least one transition occurs on $xy \times [0,\beta]$, and $\eta_{xy}=0$ otherwise. The percolation parameter is $p = \bbP(\eta_{xy}=1) = 1 - \e{-\beta}$. Let $\caC(\omega)$ denote the set of percolation clusters of $\eta(\omega)$. One can check that each loop $\gamma \in \caL(\omega)$ is contained in some cluster $C \in \caC(\omega)$ in the sense that $\gamma \subset C \times [0,\beta]$. Then $|\caC(\omega)| \leq |\caL(\omega)|$ and the presence of long loops is possible only when percolation occurs.

The critical parameter for the cubic lattice is $p_{\rm c} = 0.2488$, which gives $\beta_{\rm c}^{\rm per} = -\log(1-p_{\rm c}) = 0.286$. We have indeed found that $\beta_{\rm c}(u) > \beta_{\rm c}^{\rm per}$ for all $u \in [0,1]$, and the inequality is strict.

\medskip\noindent
{\bf B. Random cluster and Ising model.}
The random cluster model is similar to bond percolation, but with configurations $\eta$ receiving the extra weight $q^{|\caC(\omega)|}$. The random cluster model is closely related to the $q$ state Potts model; their transition parameters satisfy $p_{\rm c}^{{\rm r.c.} (q)} = 1 - \e{-\beta_{\rm c}^{{\rm Potts} (q)}}$. The case $q=2$ is equivalent to the Ising model with $2\beta_{\rm c}^{\rm Ising} = \beta_{\rm c}^{{\rm Potts} \, (q)}$. See \cite[Section 3.8]{FV} for an excellent introduction to this topic. The Curie temperature for the three-dimensional Ising model is $T_{\rm c} = 4.51$, which allows to deduce that $\beta_{\rm c}^{{\rm r.c.} (q=2)} = \beta_{\rm c}^{{\rm Potts} \, (q=2)} = 0.443$.

There is no natural comparison between $\beta_{\rm c}^{{\rm r.c.} \, (q=2)}$ of the random cluster model, and $\beta_{\rm c}(u)$ of the random loop model. But it can be compared with $\beta_{\rm c}^{\rm per}$ and $\beta_{\rm c}^{(2)}$, see \eqref{comparison}.

\medskip\noindent
{\bf C. Quantum Heisenberg models.}
Let $S^{1}, S^{2}, S^{3}$ denote the usual spin operators in $\bbC^{2}$ that satisfy $[S^{1},S^{2}] = \ii S_{3}$ and further cyclic relations. Consider the quantum Hamiltonian that acts on the Hilbert space $\otimes_{x \in \Lambda} \bbC^{2}$:
\be
H_{\Lambda} = -2 \sumtwo{\{x,y\} \subset \Lambda}{\|x-y\|=1} \Bigl( S_{x}^{1} S_{y}^{1} + (2u-1) S_{x}^{2} S_{y}^{2} + S_{x}^{3} S_{y}^{3} \Bigr).
\ee
The case $u=1$ corresponds to the spin $\frac12$ Heisenberg ferromagnet; the case $u=\frac12$ is the quantum XY model; the case $u=0$ is unitarily equivalent to the Heisenberg antiferromagnet, provided $\Lambda$ is bipartite.

The partition function and the quantum correlations can be expressed using random loop models where realisations $\omega$ receive the extra weight $2^{|\caL(\omega)|}$ \cite{Toth,AN,Uel}. In particular, $\beta_{\rm c}^{(2)}(u)$ is equal to the inverse Curie temperature of the quantum model. Numerical studies have found that $\beta_{\rm c}^{(2)}(1) = 0.59$ \cite{TAW} and $\beta_{\rm c}^{(2)}(0) = 0.53$ \cite{San}.

The extra weight encourages the system to have more, smaller loops, so we can expect $\beta_{\rm c}^{(2)}(u) \geq \beta_{\rm c}(u)$ for all $u$. Besides, we observe that $\beta_{\rm c}^{(2)}(0) < \beta_{\rm c}^{(2)}(1)$, as in the absence of the weight.

Regarding the quantum XY model, Stefan Wessel has just performed numerical calculations and he obtained the value $T_{\rm c} = 1.008 \pm 0.001$ for the Hamiltonian with interaction $-\sum_{\{x,y\}} (S_{x}^{1} S_{y}^{1} + S_{x}^{3} S_{y}^{3})$ \cite{Wes}. This implies that $\beta_{\rm c}^{(2)}(\frac12) \approx 0.496$. We conjecture that $\beta_{\rm c}^{(2)}(u)$ has a shape similar to that of $\beta_{\rm c}(u)$.

Let us summarise the discussion above with the following inequalities; for all $u \in [0,1]$,
\be
\label{comparison}
\beta_{\rm c}^{\rm per} = 0.286 \leq \left\{ \begin{matrix} \beta_{\rm c}(u) \in [0.313,0.361] \\ \\ \beta_{\rm c}^{{\rm r.c.} \, (q=2)} = 0.443 \end{matrix} \right\} \leq \beta_{\rm c}^{(2)}(u) \in [0.496,0.59]. 
\ee

\section{Joint distribution of the lengths of long loops}
\label{sec PD}

\subsection{Calculation of the moments of Poisson-Dirichlet}

We check the presence of the Poisson-Dirichlet distribution by looking at its moments, following \cite{NCSOS}. Let $n_{1} \geq \dots \geq n_{k}$ be integers. The calculation of the moments can be achieved by starting from another representation of the Poisson-Dirichlet distribution due to Kingman \cite{Kin}. Let $Z_{1}, \dots, Z_{N}$ be i.i.d.\ random variables with Gamma$(\frac\vartheta N)$ distribution (that is, their probability density function is $s^{\frac\vartheta N - 1} \e{-s} / \Gamma(\frac\vartheta N)$ for $0 \leq s < \infty$). Let $S = Z_{1} + \dots + Z_{N}$. Consider the sequence
\be
\Bigl( \frac{Z_{1}}S, \dots, \frac{Z_{N}}S \Bigr)
\ee
and reorder it in decreasing order, so it forms a random partition of $[0,1]$. As $N\to\infty$, this partition turns out to converge to PD$_{[0,1]}(\vartheta)$. The following two observations are keys to our calculations:
\begin{itemize}
\item $S$ is a Gamma$(\vartheta)$ random variable;
\item $S$ is independent of $(\frac{Z_{1}}S, \dots, \frac{Z_{N}}S)$.
\end{itemize}
For given integers $n_{1}, \dots, n_{k} \geq 0$, using the independence of $S$ from the partition, we have
\be
\begin{split}
\bbE_{{\rm PD}_{[0,1]}(\vartheta)} \Bigl( \sumtwo{j_{1}, \dots, j_{k} \geq 1}{\rm distinct} Y_{j_{1}}^{n_{1}} \dots Y_{j_{k}}^{n_{k}} \Bigr) &= \lim_{N\to\infty} \frac{N!}{(N-k)!} \; \bbE \Bigl( \Bigl( \frac{Z_{1}}S \Bigr)^{n_{1}} \dots \Bigl( \frac{Z_{k}}S \Bigr)^{n_{k}} \Bigr) \\
&= \lim_{N\to\infty} \frac{N!}{(N-k)!} \; \frac{\bbE\bigl( S^{n_{1} + \dots + n_{k}} (\frac{Z_{1}}S)^{n_{1}} \dots (\frac{Z_{k}}S)^{n_{k}} \bigr)}{\bbE( S^{n_{1} + \dots + n_{k}})} \\
&= \lim_{N\to\infty} \frac{N!}{(N-k)!} \; \frac{\Gamma(\vartheta) \, \bbE \bigl( Z_{1}^{n_{1}} \dots Z_{k}^{n_{k}} \bigr)}{\Gamma(\vartheta + n_{1}+\dots+n_{k})}.
\end{split}
\ee
We also used $\bbE(S^{a}) = \Gamma(\vartheta+a) / \Gamma(\vartheta)$. Since the $Z_{i}$s are independent,
\be
\bbE \bigl( Z_{1}^{n_{1}} \dots Z_{k}^{n_{k}} \bigr) = \prod_{i=1}^{k} \frac{\Gamma(\vartheta/N + n_{i})}{\Gamma(\vartheta/N)}.
\ee
Recall that $\Gamma(\vartheta/N) \sim N/\vartheta$ as $N\to\infty$, so that $\frac{N!}{(N-k)! \Gamma(\vartheta/N)^{k}} \to \vartheta^{k}$. We obtain
\be
\label{c'est important !}
\bbE_{{\rm PD}_{[0,1]}(\vartheta)} \Bigl( \sumtwo{j_{1}, \dots, j_{k} \geq 1}{\rm distinct} Y_{j_{1}}^{n_{1}} \dots Y_{j_{k}}^{n_{k}} \Bigr) = \frac{\vartheta^{k} \, \Gamma(\vartheta) \, \Gamma(n_{1}) \dots \Gamma(n_{k})}{\Gamma(\vartheta + n_{1} + \dots + n_{k})}.
\ee
This important formula appears in \cite{NCSOS}. Its derivation there is different; it involves another loop soup model, assumes the presence of Poisson-Dirichlet, and uses a ``supersymmetry'' method.

Eq.\ \eqref{c'est important !} holds for Poisson-Dirichlet on the interval $[0,1]$. The formula for the interval $[0,m]$ is identical, except for the additional factor $m^{n_{1}+\dots+n_{k}}$. Combining Conjecture \ref{conj2} and Eq.\ \eqref{c'est important !}, we get the following exact formula.

\begin{conjecture}
\label{conj3}
The moments of the lengths of the loops are given by
\[
\bbE \biggl( \sumtwo{j_{1}, \dots, j_{k} \geq 1}{\rm distinct} \Bigl( \frac{\ell_{j_{1}}}{|\Lambda|} \Bigr)^{n_{1}} \dots \Bigl( \frac{\ell_{j_{k}}}{|\Lambda|} \Bigr)^{n_{k}} \biggr) = m(\beta)^{n_{1}+ \dots + n_{k}} \frac{\vartheta^{k} \, \Gamma(\vartheta) \, \Gamma(n_{1}) \dots \Gamma(n_{k})}{\Gamma(\vartheta + n_{1} + \dots + n_{k})},
\]
where $\vartheta=1$ for $u=0$ or $u=1$, and $\vartheta=\frac12$ for $0<u<1$.
\end{conjecture}

With minor modifications, this formula also applies to a wide range of ``loop soup'' models. And indeed, it was derived in \cite{NCSOS} in the context of O(N) loop models.

\subsection{Numerical results}

We now calculate numerically some moments of the joint distribution of the loops and compare the results with Conjecture \ref{conj3}. First, moments involving a single loop. Let $m_{n_{1}}(\beta)$ be the value of $m(\beta)$ in Conjecture \ref{conj3} when $n_{1} \in \bbN$ and $n_{j}=0$ for $j\geq2$. Explicitly, we have
\be
m_{n_{1}}(\beta) = \biggl[ \frac{\Gamma(\vartheta+n_{1})}{\vartheta \Gamma(\vartheta) \Gamma(n_{1})} \bbE \Bigl( \sum_{j\geq1} \Bigl( \frac{\ell_{j}}{|\Lambda|} \Bigr)^{n_{1}} \Bigr) \biggr]^{1/n_{1}}.
\ee
Our numerical results deal with $\beta=1$, $u \in \{0, \frac12, 1\}$, and $n_{1} \in \{2,3,4,5\}$ and they are listed in Table \ref{table moments0}. They show that $m_{n_{1}}(\beta)$ is quite constant in $n_{1}$, apart from numerical fluctuations and finite-size corrections. This confirms Conjecture \ref{conj3}. Notice that it confirms in particular that $\vartheta$ is either 1 or $\frac12$, depending on the value of $u$.
\begin{table}[htb]
\begin{tabular}{lc|ccc}
&& \multicolumn{3}{c}{$u$} \\
&& 0 & $\frac12$ & 1 \\
\hline
\multirow{4}{*}{$n_{1}$} & 2 & 0.8925 & 0.9585 & 0.9310 \\
 & 3 & 0.8968 & 0.9587 & 0.9276 \\
 & 4 & 0.8815 & 0.9595 & 0.9217 \\
 & 5 & 0.8930 & 0.9528 & 0.9356 \\
\end{tabular}
\medskip
\caption{Numerical values of $m_{n_{1}}(\beta)$ for $\beta = 1$, and $L =160$.}
\label{table moments0}
\end{table}

Next, let $m_{n_{1},n_{2}}(\beta)$ be as $m_{n_{1}}(\beta)$, but with $n_{2} \neq 0$. The expression is
\be
\label{def moments}
m_{n_{1},n_{2}}(\beta) = \biggl[ \frac{\Gamma(\vartheta + n_{1} + n_{2})}{\vartheta^{2} \Gamma(\vartheta) \Gamma(n_{1}) \Gamma(n_{2})} \; \bbE \Bigl( \sumtwo{j_{1},j_{2}\geq1}{\rm distinct} \Bigl( \frac{\ell_{j_{1}}}{|\Lambda|} \Bigr)^{n_{1}} \Bigl( \frac{\ell_{j_{2}}}{|\Lambda|} \Bigr)^{n_{2}} \Bigr) \biggr]^{1 / (n_{1}+n_{2})}.
\ee
The numerical values are listed in Tables \ref{table moments1}--\ref{table moments3} for $u=0,\frac12,1$. Notice that we avoid the values $n_{i}=1$ because of undesirable effects due to small loops.

\begin{table}[htb]
\begin{tabular}{lc|cccc}
&& \multicolumn{4}{c}{$n_{2}$} \\
&& 2 & 3 & 4 & 5 \\
\hline
\multirow{4}{*}{$n_{1}$} & 2 & 0.9031 & 0.8924 & 0.8915 & 0.9001 \\
 & 3 & . & 0.9056 & 0.9001 & 0.8985 \\
 & 4 & . & . & 0.8987 & 0.8947 \\
 & 5 & . & . & . & 0.8946 \\
\end{tabular}
\medskip
\caption{Numerical values of $m_{n_{1},n_{2}}(\beta)$ for $u=0$, $\beta = 1$, and $L =160$. Here, $m_{L}(\beta) = 0.872$.}
\label{table moments1}
\end{table}

\begin{table}[htb]
\begin{tabular}{lc|cccc}
&& \multicolumn{4}{c}{$n_{2}$} \\
&& 2 & 3 & 4 & 5 \\
\hline
\multirow{4}{*}{$n_{1}$} & 2 & 0.9576 & 0.9579 & 0.9562 & 0.9593 \\
 & 3 & . & 0.9671 & 0.9600 & 0.9580 \\
 & 4 & . & . & 0.9539 & 0.9479 \\
 & 5 & . & . & . & 0.9523 \\
\end{tabular}
\medskip
\caption{Numerical values of $m_{n_{1},n_{2}}(\beta)$ for $u=0.5$, $\beta = 1$, and $L =160$. Here, $m_{L}(\beta) = 0.949$.}
\label{table moments2}
\end{table}

\begin{table}[htb]
\begin{tabular}{lc|cccc}
&& \multicolumn{4}{c}{$n_{2}$} \\
&& 2 & 3 & 4 & 5 \\
\hline
\multirow{4}{*}{$n_{1}$} & 2 & 0.9286 & 0.9295 & 0.9318 & 0.9315 \\
 & 3 & . & 0.9350 & 0.9256 & 0.9341 \\
 & 4 & . & . & 0.9349 & 0.9282 \\
 & 5 & . & . & . & 0.9210 \\
\end{tabular}
\medskip
\caption{Numerical values of $m_{n_{1},n_{2}}(\beta)$ for $u=1$, $\beta = 1$, and $L =160$. Here, $m_{L}(\beta) = 0.925$.}
\label{table moments3}
\end{table}

We observe that $m_{n_{1},n_{2}}(\beta)$ does not depend much of $n_{1},n_{2}$, as expected from Conjecture \ref{conj3}. This also guarantees that the value of $\vartheta$ has been conjectured correctly. Variations in the values of $m_{n_{1},n_{2}}(\beta)$ can be dismissed  as random fluctuations and finite-size effects.

To summarise, we have studied numerically the joint distribution of the lengths of macroscopic loops in a family of loop models in three dimensions. These loop models are motivated by their close relations to certain quantum spin systems and include in particular the random interchange model. We have observed the presence of the Poisson-Dirichlet distribution with parameters $\vartheta = 1$ and $\frac12$ as conjectured in \cite{GUW,Uel}.

\medskip
\noindent
{\bf Acknowledgments: } We are grateful to Stefan Wessel for sending us the result of his numerical calculation of the critical temperature of the quantum XY model.

\end{document}